\documentstyle[prl,aps,preprint]{revtex}
\begin{document}

\def\be{\begin{equation}}
\def\ee{\end{equation}}
\def\vr{{\bf r}}
\def\vz{{\bf z}}
\def\vs{{\bf s}}
\def\cs{\vec{\cal S}}
\def\gamm{\gamma_{\perp}}
\def\gp{\gamma_{\parallel}}

\draft
\title{\bf Low-Magnetic Field Critical Behavior
in Strongly Type-II Superconductors}
\author{
Zlatko Te\v sanovi\' c
}
\address{
Department of Physics and Astronomy,
The Johns Hopkins University,
Baltimore, MD 21218}

\maketitle
\begin{abstract}
A new description is proposed
for the low-field critical
behavior of type-II superconductors.
The starting point is the Ginzburg-Landau theory
in presence of an external magnetic field ${\bf H}$.
A set of fictitious vortex variables and
a singular gauge transformation are used to
rewrite a finite ${\bf H}$
Ginzburg-Landau functional in terms of
a complex scalar field of {\it zero} average
vorticity. The continuum limit of the
transformed problem takes the form of
an ${\bf H}=0$ Ginzburg-Landau
functional for a charged field
coupled to a fictitious
`gauge' potential which arises from long wavelength
fluctuations in the background liquid of field-induced
vorticity.  A possibility of a
novel phase transition involving zero vorticity
degrees of freedom and formation of a {\it uniform}
condensate is suggested.
A similarity to the superconducting [Higgs]
electrodynamics and the nematic-smectic-A
transition in liquid crystals is noted.
The experimental situation is discussed.
\end{abstract}

\narrowtext

There are two basic
theoretical approaches to the
fluctuation behavior of high temperature superconductors (HTS) and
other strongly type-II systems in a magnetic
field: The Ginzburg-Landau (GL)
theory confined to the lowest
Landau level (LLL) for Cooper pairs
(the GL-LLL theory)\cite{gl-lll}
and the XY-model approach, in which one
focuses on the field-induced London vortices and
suppresses other superconducting fluctuations.\cite{vortex}
Generally, it is expected that the GL-LLL theory describes the high-field
regime, close to $H_{c2}(T)$, behavior,
while the London vortex theory should be
appropriate at low fields, far from $H_{c2}(T)$. It is
often presumed that there is a smooth crossover between
the two regimes.

It is argued here that there is an important
physical difference between the
critical behaviors at high and low fields, reflected
in the nature of low energy fluctuations. At high
fields, all such fluctuations are
represented by the motion of $N_{\phi}$ field-induced vortices.
This is so because the states within the LLL can be expressed
as different configurations of zeros of the holomorphic
order parameter \cite{zt}. Higher LLs have a finite gap and contribute
only by renormalizing various terms in the GL-LLL theory.
In contrast, at low fields, even if we imagine that the
field-induced vortices are fixed in their positions, there
are low energy fluctuations of
many degrees of freedom which are not associated with
field-induced vortices: non-singular phase fluctuations,
vortex-antivortex pairs, vortex loops, etc. These fluctuations
originate from linear combinations of many LLs.
It is precisely these {\it zero vorticity}
degrees of freedom that produce the
zero-field superconducting transition. At finite but low fields,
it is reasonable to expect that these degrees of freedom
still account for most of the entropy change in the critical
regime and dominate various thermodynamic functions.

In this paper a new description is derived
of the low-field critical behavior
associated with these {\it zero vorticity} fluctuations. The
key step is to rewrite the original partition function
of the GL theory in terms of
a complex scalar field $\Phi$ of {\it zero average vorticity},
instead of the usual superconducting order
parameter field $\Psi$ whose
average vorticity is $N_{\phi}$, as fixed by the magnetic
field. This vorticity shift by $N_{\phi}$ in the functional space of
the fluctuating order parameter is
accomplished by first introducing a set of $N_{\phi}$ auxiliary
vortex variables, which we may call `shadow' vortices (or svortices),
and then performing a `singular' gauge transformation $\Psi\to\Phi$.
The continuum hydrodynamic limit of the transformed problem is
a field theory involving a complex scalar field $\Phi$
in an effective average magnetic field equal to {\it zero}
coupled to a {\it fictitious} `gauge'
potential ${\cs}$ produced by local
fluctuations of svorticity around its average value. In
intuitive terms, $\Phi$
represents those degrees of freedom of the original
superconducting order parameter which cannot be
reduced to motion of field-induced vortices, while
${\cs}$ arises from the
{\it long wavelength} density and current
fluctuations in the background system of these field-induced
London vortices. The physical insight gained by this transformation
is that we have now uncovered in the original GL problem
those hidden off-diagonal correlations, represented by $\Phi$, whose
range extends {\it far beyond} the average separation between
field-induced vortices (set by the magnetic length).
At the same time, the range of {\it original}
superconducting correlator involving field $\Psi$ remains
{\it limited} by the magnetic length. It is these novel
off-diagonal correlations with a range longer
than the magnetic length that control the low-field critical behavior.
Furthermore, a possibility is pointed out of
a novel {\it finite-field} [FF] phase transition involving divergence
of a certain susceptibility related to $\Phi$ and ${\cs}$ and
formation of an unusual {\it uniform} condensate.
As external field tends to zero the line of the FF phase transitions
terminates in the familiar zero-field [ZF] critical point at
$T=T_{c0}$. Next, a connection is made
between the FF critical fluctuations
and the critical behavior of
liquid crystals and the ordinary superconducting [Higgs]
electrodynamics at zero field. An intuitive picture
is proposed for this FF transition in the context
of the 3D XY-model where the $\Phi$-ordered phase
is identified as the {\it vorticity-incompressible} London
liquid state while the $\Phi$-disordered
state corresponds to the high-temperature {\it compressible}
fluid of unbound vortex loops. Finally, the
experimental situation is discussed.

The anisotropic GL partition function
is the appropriate starting point for HTS and other
layered superconductors:
$ Z = \int {\cal D}[\Psi (\vr,\zeta)]
\exp\{- F_{GL}[\Psi (\vr,\zeta)]/T\}$,
where
\be
\label{ei}
F_{GL} =
\int d^2rd\zeta \{ \alpha \vert
\Psi \vert ^2 + {{\beta} \over {2}}
\vert \Psi \vert ^4
+\gamm \vert [{\bf\nabla_{\perp}}
+ {{2ei} \over{c}} {\bf A}]
\Psi\vert ^2 +
\gp\vert \partial_{\zeta}\Psi\vert ^2 \}~~,
\ee
$\vr = (x,y)$ and $\zeta$ are the coordinates perpendicular
and along ${\bf H}$,
${\bf \nabla}\equiv [{\bf \nabla_{\perp}},\partial_{\zeta}]$,
$\alpha =\alpha_0[(T/T_{c0})-1]$,
$\beta$, $\gamm$, $\gp$ are the GL coefficients,
and $\nabla\times{\bf A} = {\bf H}$, with the magnetic field
${\bf H}$ perpendicular to the layers.
Fluctuations of the electromagnetic field are neglected
throughout the paper ($\kappa\gg 1$).

An important feature of $F_{GL}$
is the formation of Landau
levels (LLs) for Cooper pairs.
The LL structure arises from the
quadratic part of $F_{GL}$.
The quartic interaction term, however,
mixes different LLs and acts
to suppress the LL structure in
the fluctuation spectrum. It is instructive
to divide the effect of the
quartic term into the intra-LL
and inter-LL correlations. At high fields,
$H\gg H_b$, where the
cyclotron gap between LLs is much larger
than the interaction term, only the {\it intra}-LL
correlations are important and the LL structure
will be reflected in the theory. In this regime
the GL-LLL description captures the essential
features of the physics. In the opposite limit of
low fields, $H\ll H_b$, the {\it inter}-LL
correlations become dominant and
the LL structure is suppressed
at long wavelengths. The crossover field $H_b$,
which separates the high-field LL regime from the low-field
`semiclassical' one, is given by
$H_b\sim (\theta/16)(T/T_{c0})H_{c2}(0)$,
where $\theta$ is the Ginzburg fluctuation number.
In terms of the parameters of the GL theory
$\theta\cong 2\beta H'_{c2}T_{c0}^2/
\phi_0\alpha_0^{3/2}\gp^{1/2}$, where
$H'_{c2} = [dH_{c2}/dT]$ at $T=T_{c0}$.
This expression for $H_b$ was derived in
Ref. \onlinecite{ta}\ by comparing the strength
of quartic correlations
in (\ref{ei}) with the cyclotron gap between LLs.
In HTS $\theta \sim 0.01-0.05$
and $H_b\sim 0.1 - 1$ Tesla.\cite{ta}

In the low-field limit, $H\ll H_b$, a semiclassical
description becomes possible. The idea
proposed here is to
shift the overall vorticity of the fluctuation
spectrum by $N_{\phi}$, where $N_{\phi}=
\Omega /2\pi\ell^2$, $\Omega$
and $\ell =\sqrt{c/2eH}\equiv\sqrt{\phi_0/2\pi H}$ being the
area of a layer and the magnetic length, respectively.
This shift is accomplished by
introducing a set of
$N_{\phi}$ auxiliary variables, $\{ \vz _{i}(\zeta)\}$,
which one may call `shadow' vortices (svortices).
We now change variables in the
functional integral: $\Psi (\vr,\zeta)
\to \Phi (\vr,\zeta)=\Psi (\vr,\zeta)
\exp [-i\Theta (\vr,\zeta)]$, where $\Theta (\vr,\zeta)=
\sum_i\tan ^{-1}[(y-y_{i}(\zeta))/(x-x_{i}(\zeta))]$,
$\vz _i(\zeta)=[x_i(\zeta),y_i(\zeta)]$,
and rewrite the GL functional (\ref{ei}) as:
\be
\label{eii}
\int d^2rd\zeta \{ \alpha \vert
\Phi \vert ^2 + {{\beta} \over {2}}
\vert \Phi \vert ^4
+\gamm \vert [{\bf\nabla_{\perp} }
+ i\nabla_{\perp}\Theta + {{2ei} \over{c}} {\bf A}]
\Phi \vert ^2
+\gp\vert [\partial_{\zeta}
+ i\partial_{\zeta}\Theta]\Phi \vert ^2 \}~~,
\ee
while simultaneously introducing the functional integral
over svortex variables,
$\int\prod_{i\zeta}d \vz_{i}(\zeta){\cal J}[\Phi,\{ \vz_{i}(\zeta)\}]/N!$,
in the measure.  ${\cal J}[\Phi,\{ \vz_{i}(\zeta)\}]$
is the Jacobian of this [singular] gauge transformation
and it ensures that
in the low temperature limit, where
$\Phi (\vr,\zeta)$ takes the form which minimizes
$F_{GL}$ for a given configuration of $\{ \vz_i(\zeta)\}$,
the above formulation becomes equivalent
to the standard London description.
Note, however, that svortices are fictitious
objects and do {\it not} directly correspond to the physical vortex
excitations of $\Psi (\vr,\zeta)$.
The whole transformation
is an identity if the short wavelength
behavior is properly regularized, for example, by having
$\Psi (\vr,\zeta)$ and svortices defined on dual sublattices.

The integration over svortex variables
would now lead to a new, and complicated, representation
of the original problem (\ref{ei}).
Such an exact integration is beyond reach. What has been gained,
however, is that we now have a reformulation of the
problem in terms of field $\Phi$ whose average vorticity
equals zero--Therefore, certain aspects of the
physics become more visible.
In particular,
the long-wavelength
functional for $\Phi$
can be extracted by
the following `smoothing out' procedure: New
variables are introduced,
$\rho (\vr,\zeta)\equiv
\sum_i\delta (\vr -\vz_{i}(\zeta))$,
${\bf j} (\vr,\zeta)\equiv
\sum_i[d{\bf z}_i/d\zeta]\delta (\vr -\vz _{i}(\zeta))$,
which represent microscopic
svortex density and  `current',
respectively. In terms of $[\rho,{\bf j}]$
we have
$[\nabla_{\perp}\Theta,\partial_{\zeta}\Theta]
\equiv \int d\vr' (\vr -\vr')\times [{\bf e_z}\rho (\vr',\zeta),
-{\bf j}(\vr',\zeta)]/
\vert \vr -\vr'\vert^2$,
where ${\bf e_z}$ is a unit vector along ${\bf H}$.
The GL partition function is transformed
into:
$$Z=\int {\cal D}[\Phi]
{\cal D}[\rho]
{\cal D}[{\bf j}]
\exp\{-F'_{GL}[\Phi,\rho,{\bf j}]/T
+W[\rho,{\bf j}]\}~~,~ \exp\{W[\rho,{\bf j}]\}\equiv
\int\prod_{i\zeta}\frac{d\vz _{i}(\zeta)}{N!}{\cal J}\times
$$
\be
\label{eiii}
\times\prod_{\vr}\delta\left[\rho (\vr,\zeta)-\sum_i\delta (\vr
-\vz_{i}(\zeta))\right]
\delta\left[{\bf j}(\vr,\zeta)-\sum_i\frac{d{\bf z}_i}{d\zeta }
\delta (\vr -\vz_{i}(\zeta))\right] ~~,
\ee
where $F'_{GL}$ is given by Eq. (\ref{eii}) with
$[\nabla_{\perp}\Theta,\partial_{\zeta}
\Theta ]$ expressed via $[\rho,{\bf j}]$.

The above expression is formally exact but useless, since
$\rho$ and ${\bf j}$ are wildly varying functions.
For $H\ll H_b$, however, it is expected that
replacing $\rho (\vr,\zeta)\to{\bar\rho}(\zeta) +
\delta\rho (\vr,\zeta)$ in
(\ref{eiii}),
where ${\bar\rho}(\zeta) = (2\pi\ell^2)^{-1}$ is the
average svortex density
and $\delta\rho (\vr,\zeta)$ is a
{\it smooth} function describing variations around
the average, should be adequate at wavelengths long
compared to $\ell$. Similar `smoothing' out is
performed in ${\bf j}(\vr,\zeta)$.
This leads to the `hydrodynamic'
limit of the GL functional:
$$Z=\int {\cal D}[\Phi (\vr,\zeta)]
{\cal D}[\delta\rho (\vr,\zeta)]
{\cal D}[{\bf j} (\vr,\zeta)]
\exp\{-F''_{GL}[\Phi (\vr,\zeta),\delta\rho (\vr,\zeta),{\bf j}(\vr,\zeta)]/T
+W[\delta\rho,{\bf j}]\}~~,$$
\be
\label{eiv}
F''_{GL}=
\int d^2rd\zeta \{ \alpha \vert
\Phi \vert ^2 + {{\beta} \over {2}}
\vert \Phi \vert ^4
+\gamm \vert [{\bf\nabla_{\perp}}
+ i{\vs}(\vr,\zeta)]
\Phi \vert ^2
+\gp \vert [\partial_{\zeta}
+ is_{\zeta}(\vr,\zeta)]\Phi \vert ^2 \}~~,
\ee
with the form of $W[\delta\rho,{\bf j}]$ apparent
from (\ref{eiii}) and a
new vector field
${\cs}\equiv[\vs,s_{\zeta}] =
\int d\vr' (\vr -\vr')\times [{\bf e_z}\delta\rho (\vr',\zeta),
-{\bf j}(\vr',\zeta)]/
\vert \vr -\vr'\vert^2$.

The main feature of $F''_{GL}$ is that ${\bf H}$
has now disappeared from the problem. The
vector potential ${\bf A}$ has been
canceled by the average svortex density, ${\bar\rho}$, which appears
in $\langle\nabla_{\perp}\Theta (\vr,\zeta)\rangle$ ($\langle\cdots\rangle$
denotes thermal average). The average magnetic field
felt by $\Phi$ is zero. We have thus
arrived at the following simple description:
The critical behavior of
the GL partition function (\ref{ei}) can be represented
by a complex scalar field $\Phi$ of zero average vorticity coupled
to a fictitious `sgauge' potential ${\cs}$
produced by local
fluctuations of vorticity around
its average value, $N_{\phi}$,
set by the external field.

While the above picture appears intuitive it
is by no means obviously justified. The above
derivation relies on the
{\it assumption} of separation
of long wavelength fluctuations in
$\Phi$ from the rapidly changing short
wavelength variations of the {\it microscopic},
i.e. not `smoothed out', svortex density.
To demonstrate that such separation indeed takes place
in (\ref{ei}) is no trivial task.\cite{ikeda}
In fact, the high-field, LL regime
is entirely dominated by such core effects.
In the low-field
limit of the XY-model, where
the microscopic core size,
$a$, satisfies $a\ll\ell$,
it is hoped that these core effects
eventually become irrelevant for the
long wavelength behavior.
For the rest of the paper it is
assumed that the magnetic field
is sufficiently low so that core effects in (\ref{ei}),
even if relevant, affect critical
behavior only at distances much
too long to be of practical interest. The range of
the validity of this semiclassical description
can then be established
empirically.

These clarifications noted,
I now demonstrate
the utility of this description of low-field
critical behavior. The free energy of Eq. (\ref{eiv})
can be evaluated in the mean-field approximation
$\Phi (\vr,\zeta)=\Phi_0$:
\be
\label{ev}
F_{\rm mf}[\Phi_0] = \alpha\vert\Phi_0\vert^2 +
\frac{\bar\beta}{2}\vert\Phi_0\vert^4
+\frac{\gamm}{2\pi\ell^2}\vert\Phi_0\vert^2 f_C (\Gamma)~~,
\ee
where ${\bar\beta}=\beta - 2C\gp^2\phi_0/a_{\parallel}^4
T_{c0}H$, $C$ is of order unity and
$a_{\parallel}={\rm max}(\xi_{\parallel},d)$,
with $\xi_{\parallel}$ and $d$ being the coherence
length along ${\bf H}$ and the interlayer separation,
respectively.
In deriving (\ref{ev}) I assumed
$2C\gp^2\phi_0/a_{\parallel}^4\beta T_{c0}H\ll 1$, which
is appropriate for BSCCO HTS.
$Q^2 f_C (\Gamma)$, with $\Gamma = Q^2/T \equiv
\vert\Phi_0\vert^2\gamm\xi_{\parallel}/T$,
is the free energy of the
2D one-component Coulomb
plasma (OCP), $f_C = (1/2)\ln\left(\sqrt{2}\ell /a\right)
+ g(\Gamma)$.\cite{ocp} An approximate form, reliable for
$1\agt\Gamma$, is
$$g(\Gamma)=-\frac{1}{4}\left\{2E +\ln\left[
\frac{\Gamma}{\Gamma +2}\left(\frac{2}{\Gamma +2}\right)^{2/\Gamma}
\right]\right\}~~,$$
where $E=0.5772...$ is Euler's constant.
$\Phi_0$ is determined by minimizing $F_{\rm mf}$.
Below some $T_{\Phi}(H)$, which has to be determined
numerically,  the minimum of $F_{\rm mf}$
shifts from $\Phi_0 =0$ to finite $\Phi_0$.
A simple approximate formula is
\be
\label{tphi}
T_{\Phi}(H)\cong
T_{c0}-c_1\frac{4\pi\gamm T_{c0}}{\alpha_0\phi_0}H\left\{
\frac{1-E}{2}+\frac{1}{4}\ln\left[
\frac{2c_2{\bar\beta}T_{c0}\phi_0^2}
{\pi a^2a_{\parallel}(4\pi\gamm)^2H^2}\right]\right\}~~,
\ee
where $c_{1,2}$ are constants
of order unity.  The transition is weakly
first order due to long-range interactions in OCP,
with the jump
$\Delta\Phi_0\cong\sqrt{4\pi\gamm H/2{\bar\beta}\phi_0}$.

$T_{\Phi}(H)$ [or $H_{\Phi}(T)$, as in Fig. 1]
could be interpreted
as the `fluctuation
renormalized' $H_{c2}(T)$.
This $\Phi$-transition is driven by
the growth of a novel off-diagonal order associated with
$\Phi$, and {\it not} the original
superconducting field $\Psi$. Above
the mean-field $H_{c2}(T)$, the correlation length of
$\chi_{\Phi}(\vr,\zeta)\equiv
\langle \Phi (\vr,\zeta)\Phi^*(0,0)\rangle$
in the xy-plane, $\xi_{\Phi}$,
is much shorter than $\ell$ and
is approximately equal to $\xi_{\Psi}$, the latter
being the usual superconducting correlation length,
associated with
$\chi_{\Psi}(\vr,\zeta)\equiv
\langle \Psi (\vr,\zeta)\Psi^*(0,0)\rangle$ (the `XY'
region in the inset of Fig. 1).
In the critical region below $H_{c2}(T)$,
$\xi_{\Psi}$ saturates at $\sim\ell$, but
$\xi_{\Phi}$ grows rapidly and becomes
$\gg\ell$ (the `$\Phi$' region in the inset of Fig. 1).
This illustrates the main physical idea of
this paper: While the
original pairing correlations in the critical region
remain limited by the motion of
field-induced vortices, which form a liquid both above
and below $H_{\Phi}(T)$,
there are other off-diagonal correlations in (\ref{ei})
whose range greatly exceeds $\ell$. $H_{\Phi}(T)$ has no analogue
in the high-field regime. There $H_{c2}(T)$ is only a
smooth crossover.

The mean-field theory predicts that these $\Phi$-correlations
become long-ranged below $H_{\Phi}(T)$.
To examine this prediction we
consider fluctuations in $\Phi$.
The `smoothed out' variables $[\delta\rho,{\bf j}]$
are well-defined only on lengthscales longer
than intersvortex separation ($\sim\sqrt{2\pi}\ell$).
Let us define a cut-off, $\Lambda (T,H)\agt \sqrt{2\pi}\ell$.
The fluctuations in $\Phi$ at wavelengths
shorter than $\Lambda$ are integrated out. Finally, the functional
integral over $[\delta\rho,{\bf j}]$ is replaced by the one
over ${\cs}\equiv[{\bf s},s_{\zeta}]$. The result is:
$Z\to\int {\cal D}[\Phi]
{\cal D}[{\cs}]
\exp\{-{\cal F}_{GL}/T
+{\cal W}\}$,
$${\cal F}_{GL}=
\int d^2rd\zeta
\{\alpha \vert
\Phi \vert ^2 + {{\beta} \over {2}}
\vert \Phi \vert ^4
+\gamm \vert [{\bf\nabla_{\perp} }
+ i{\vs}(\vr,\zeta)]
\Phi \vert ^2
+\gp \vert [\partial_{\zeta}
+ is_{\zeta}(\vr,\zeta)]\Phi \vert ^2+$$
\be
\label{evi}
+K^0_{\parallel}(\nabla\times{\cs})^2_{\parallel}
+K^0_{\perp}(\nabla\times{\cs})^2_{\perp}\}~~,
\ee
where
$K^0_{\parallel,\perp} (T,H)$ is
the bare `stiffness' of the `sgauge'
field ${\cs}
\equiv
[{\bf s},s_{\zeta}]$, produced by the short
wavelength ($<\Lambda$) fluctuations in $\Phi (\vr,\zeta)$
which also renormalize the GL coefficients.
At $T\sim T_{c0}$,
$K^0_{\parallel,\perp}\sim [\sqrt{\gamm /\gp },
\sqrt{\gp /\gamm }]\times T_{c0}\Lambda$.\cite{schmid}
Higher order terms in ${\cs}$ which also are
generated by the integration of short wavelength ($<\Lambda$)
modes are presumed irrelevant. This procedure is internally consistent
in the hydrodynamic limit. Similarly,
${\cal W}[{\bf s},s_{\zeta}]$, which is simply $W$ of Eq. (\ref{eiii})
reexpressed in terms of $\cs$, is assumed to have
an expansion in $\cs (\vr,\zeta)$.
In addition, the above integration procedure with a
specified cut-off may produce
quadratic terms in $\cs$ which
violate the fictitious sgauge invariance, an example
being the mass term $m_s^2\cs^2$. All such
terms will be absorbed into the
redefinition of ${\cal W}[{\bf s},s_{\zeta}]$.
Conversely, the `sgauge invariant' part
of ${\cal W}$ will be absorbed into $K^0_{\parallel},
K^0_{\perp}$.
The remaining fluctuations in $\Phi$ and those of $\cs$
come from wavelengths $>\Lambda$.
${\cal F}_{GL}$ in (\ref{evi}) defines
the bare level functional which serves as the starting point
for study of the finite-field critical behavior. Note that, for $H\to 0$
along the $T=T_{\Phi}(H)$ line,
$K^0_{\parallel,\perp}\propto \Lambda\propto 1/\sqrt{H}\to\infty$,
the ${\cs}$ fluctuations in ${\cal F}_{GL}$ are suppressed,
and the $\Phi$-transition goes into
the zero-field superconducting transition.

${\cal F}_{GL}$ has a form reminiscent
of the standard superconductor [Higgs]
electrodynamics (SHE) at zero field (the anisotropy
is easily rescaled out\cite{blatter}). The
fine structure constant is unity rather
than 1/137, the GL parameter is
$\kappa^2_{s(\parallel,\perp)}=
\alpha_0K^0_{\perp,\parallel}(T,H)/\gamma_{\parallel,\perp}^2$,
and the gauge is fixed in an unusual way.\cite{gauge}
This `svortex gauge' has a physical origin in the
underlying connection between $[{\vs}, s_{\zeta}]$ and
$[\delta\rho,{\bf j}]$.  It is, however, awkward to work with
since the long wavelength
non-singular phase fluctuations of $\Phi$, with
$\vert\Phi\vert=\Phi_0$ fixed, still couple to ${\cs}$.
The presence of this coupling allows only power law
correlations in
$\langle\Phi (\vr,\zeta)\Phi^*(0,0)\rangle$
at low temperatures. It is expedient to introduce new
variables:
$\Phi\to {\bar\Phi}=\Phi\exp [i\Pi (\vr,\zeta)]$,
${\cs}\to {\bf S}={\cs} - \nabla\Pi$, where
$\Pi =\nabla ^{-2}\partial_{\zeta}s_{\zeta}$.
This is a simple `gauge transformation' as far as
${\cal F}_{GL}$ is concerned. This allows us to
extract the leading two-point correlations
in (\ref{evi}), which are
$\langle{\bar\Phi} (\vr,\zeta){\bar\Phi}^*(0,0)\rangle =
\langle\Phi (\vr,\zeta)\exp[i\Pi (\vr,\zeta)]
\Phi^*(0,0)\exp[-i\Pi (0,0)]\rangle$, and not
$\langle\Phi (\vr,\zeta)\Phi^*(0,0)\rangle$.
These ${\bar\Phi}$-correlations involve combined
phase fluctuations in $\Phi$ and the svortex transverse
current fluctuations which enter through $\Pi$
via $s_{\zeta}$. ${\bar\Phi}$-order
is expected to be long ranged below $T_{\Phi}(H)$.

After this `gauge transformation',
${\cal F}_{GL}$ in (\ref{evi})
becomes equivalent to SHE in the Coulomb gauge,
$\nabla\cdot {\bf S}=0$ (apart from the anisotropy).
${\cal W}[{\cs}]$ in contrast, does not possess this
fictitious gauge invariance of ${\cal F}_{GL}$.
Its form reflects
only the spatial symmetries of the original problem
(\ref{ei}). This form changes
when ${\cs}\to {\bf S}$.
A similar situation arises
in studies of the nematic-smectic-A
transition in liquid crystals.\cite{degennes}
There one also has
the part which is equivalent to SHE and additional
terms which reflect spatial symmetries of that problem.
Here I adopt the
renormalization group (RG) analysis of
the nematic-smectic transition by
Halperin, Lubensky and collaborators.\cite{lubensky}
An important result is
that the `pure' isotropic SHE is one of the fixed points of
(\ref{evi}).  At this fixed point
$K_{\parallel}, K_{\perp}\to K^*$
and ${\cal W}/K^*\to 0$.
Furthermore, this fixed point is stable
to finite ${\cal W}$ perturbations to all orders
in the $\epsilon$-expansion.
Consequently, if the bare
$K^0_{\parallel}, K^0_{\perp}$ are large enough, the
RG flows should be attracted to the SHE fixed point.
It is difficult, however,
to evaluate $K^0_{\parallel}(T,H)$ and
$K^0_{\perp}(T,H)$ from `first
principles' with a precision greater than what
is given below Eq. (\ref{evi}). At this point further
approximations become necessary.
An estimate based on the self-consistent
integration of fluctuations in $\Phi$ at wavelengths
short compared to $\Lambda$ suggests
$K^0_{\parallel},K^0_{\perp}$ that are growing exponentially
for $H\to 0$, at $T <  T_{c0}$, but this
is probably too crude. A more efficient
alternative approach might be to treat
$K^0_{\parallel},K^0_{\perp}$ given below Eq. (\ref{evi})
as a phenomenological
input to the theory and proceed to compute various
consequences of ${\cal F}_{GL}$. The comparison to experiments
and numerical simulations
can then be used to establish more precise values of
$K^0_{\parallel},K^0_{\perp}$.
At any rate, for low enough fields,
$K^0_{\parallel},K^0_{\perp}$ become large in the critical region
and the plausible scenario within the $\epsilon$-expansion is
that the fluctuation behavior would show
crossover from mean-field to anisotropic SHE
to isotropic SHE and ultimately
to a very weak first order transition at
renormalized $H_{\Phi}(T)$.\cite{first} This SHE scenario
is valid if the `mass terms' for ${\cs}$ are either absent
or small at the bare level, i.e. if
$K^0_{\parallel,\perp}/m_s^2\gg\Lambda^2$, where
$m_s$ is the `bare' mass of ${\cs}$.\cite{footxii}

The SHE phase transition has been studied also by
Dasgupta and Halperin \cite{dasgupta} using the lattice
superconductor model (LSM). They have concluded
that, within LSM, the transition appears continuous
and is in the `inverted XY' universality class,
defined by the interacting vortex loops. In our
model this implies a proliferation of unbound large
vortex loops in ${\bar\Phi}(\vr,\zeta)$ at
$H_{\Phi}(T)$. They also find that the transition
moves to lower temperature as the
stiffness of the gauge field is reduced and
completely disappears below certain
critical value of the stiffness. This should be
the case here for
higher fields, where
$K^0_{\parallel},K^0_{\perp}$ are limited by
core effects and may become large only
at comparatively low temperatures.
Ultimately, it is expected
that $T_{\Phi}(H) \to 0$ as
$H$ increases toward $H_b$. At these higher fields,
where  $K^0_{\parallel},K^0_{\perp}$ are getting smaller,
the sgauge field fluctuations are
enhanced and transition may become discontinuous
as observed in numerical simulations.\cite{bart}
This `transition' to the high-field limit,
represented by the dotted line in Fig. 1, is
beyond the semiclassical approximation of this
paper. An interesting question in this
context is the interference between the $\Phi$-transition
and the London (s)vortex liquid-solid (LVL-LVS) transition
(the crossing between the dotted $H_{\Phi}(T)$ line and the solid
LVL-LVS line in Fig. 1) which is left for future study.

The above connection to the critical behavior of
SHE is theoretically appealing and deserves
additional discussion (it is now assumed that
the bare mass, $m_s$, is negligible). The problem here is
that the critical behavior of SHE itself is still
not fully understood, as illustrated
above.  The $\epsilon$-expansion predicts the first
order transition while various numerical studies
indicate a continuous transition, at least for
strongly type-II systems.
Furthermore, the $1/N$-expansion also favors continuous
transition.\cite{leo}  Interestingly,
the problem of the finite-field [FF] critical behavior in type-II
superconductors provides additional
motivation for the study of SHE since the `fictitious' sgauge
Higgs electrodynamics introduced here has a much larger
intrinsic fine structure constant than the real SHE (unity
versus 1/137). Consequently, the
domain of FF critical fluctuations could be considerably
wider than that of the real SHE at the zero-field [ZF]
normal-superconducting transition.
While the above uncertainties concerning SHE
remain to be resolved, it is still possible to exploit this
connection to make some general remarks on
type-II superconductors in a low magnetic field,
assuming their fluctuation behavior is faithfully
represented by the 3D XY model.
The low-temperature, Meissner phase of SHE is related
to the `London (s)vortex liquid' (LVL) state
of a type-II superconductor (see Fig. 1). This is an incompressible
${\bar\Phi}$-ordered liquid phase with
long-range interactions between (s)vortices and
with the overall vorticity locked at $N_{\phi}$. Large
thermally excited vortex loops are suppressed.
The strength of this long-range London interaction
is given by the `photon' mass of
the fictitious sgauge electrodynamics which is directly
related to the `helicity modulus'
of the condensed ${\bar\Phi}$ field. Thus, as we
approach $T_{\Phi}(H)$ from below, the long-range ``London"
component of the (s)vortex interaction is renormalized
by fluctuations in the same fashion as the London magnetic
penetration depth in the {\em real} SHE.
Above $T_{\Phi}(H)$ large thermally excited
vortex loops proliferate across the sample and the system behaves
like a compressible (s)vortex fluid, with the
fluctuating overall vorticity. The
fictitious sgauge `photon' is now massless, the
${\bar\Phi}$-order is absent and (s)vorticity density-density
correlations and $\Phi$-correlations are both short-ranged. This
phase can be identified with the normal state.

I now state potential
sources of concern with the above
description. The ${\cal W}[{\cs}]$ term
could produce subtle non-perturbative
effects on the critical behavior. Furthermore,
the core effects, which have been
ignored on the basis of $a/\ell\ll 1$, might
have a non-trivial effect at long wavelengths which
could modify critical behavior and
suppress ${\bar\Phi}$-order,
even at low fields. Still, I expect the description
proposed here to retain its usefulness
at low fields near $T_{c0}$
and at the lengthscales typically
encountered in experiments. This is so because
the present theory does perform at least one important
task: it extracts from the original GL functional
those off-diagonal correlations whose range extends
well beyond the magnetic length, $\ell$. This is illustrated
in the inset of Fig. 1. Above the dashed-dotted line
the ${\bar\Phi}$-correlator has basically the same range
as the standard superconducting $\Psi$-correlator.
In this region, labeled as `XY', we can exploit the
proximity of the zero-field [ZF] critical point to
describe the physics and construct various thermodynamic
quantities. Below the dashed-dotted line, in the
region labeled as `$\Phi$', the
range of the superconducting $\Psi$-correlator
saturates at $\ell$, while ${\bar\Phi}$-correlations
continue to grow and ultimately diverge, at least
within the present description based on the
effective functional ${\cal F}_{GL}$ (\ref{evi}).
To be sure, since ${\cal F}_{GL}$ itself results from a particular
approximate way of taking the continuum limit,
one must allow for the possibility
that additional relevant terms, not included
in the present description, might modify the ultimate
long wavelength behavior of the problem.
Such subtle issues not withstanding,
the clear message of the present approach,
and the one likely to remain in place, is
that the fluctuation behavior in the `$\Phi$'  region
(Fig. 1) must be governed by the proximity
to some new {\em finite-field} [FF] critical point
and {\em not} to the zero-field transition. It
appears likely that ${\cal F}_{GL}$ captures
at least basic features of the physics associated
with this new critical behavior.

There is empirical support for the picture
presented in this paper.
$H_{c2}(T)$ determined from magnetization measurements of
Ref. \cite{ott} shows expected
linear behavior at high fields but
deviates from linearity at low fields. This
deviation is consistent with $T_{\Phi}(H)$ of Eq. (\ref{tphi})
if we relate the crossover in magnetization
to the point where $\langle |\Phi |^2\rangle$
starts growing.
A detailed analysis of
the specific heat data in 1-2-3 HTS\cite{cv} in terms of
the XY-model critical scaling leads to poor
agreement unless one allows for the
[unexplained] strong field dependence of
the coefficients.\cite{pierson} This
strong field dependence can be interpreted
as the crossover from  the `XY' to
the $\Phi$-critical behavior described
by (\ref{evi}). This crossover takes place
as one moves from the high-temperature regime,
$\xi_{\Phi}<\ell$, to the true critical
regime, $\xi_{\Phi}\gg\ell$.
Finally, Li and Teitel\cite{teitel} have reported Monte
Carlo simulations of the XY-model
which show a suppression of large vortex
loops followed by a sharp onset of the
helicity modulus for field-induced vortices.
This effect should arise
as a consequence of the ${\bar\Phi}$-ordering,
with the disappearance of large vortex loops leading
to a sharp increase in the line-tension of
field-induced vortices.
It is clear, however, that additional experimental
and computational effort will be needed before a
complete picture of the low-field critical
behavior is in place. It is hoped that the present work
will stimulate such developments.

In summary, the main advances reported in this paper can be
viewed as twofold: First, at a conceptual
level, a new description is
derived for the low-magnetic field
critical behavior of the GL theory. In contrast to the
GL theory in high fields, where the only fluctuating
degrees of freedom are positions of field-induced vortices,
the low-field critical behavior is dominated by
thermally induced zero-vorticity excitations, like
vortex-antivortex pairs, vortex loops, etc. By shifting
the vorticity in the original GL partition function
and taking the hydrodynamic limit, the low-field
critical behavior is related to the field theory
describing a complex scalar field in a zero average
magnetic field coupled to a fictitious gauge field
produced by the long wavelength fluctuations in the
background system of field-induced vorticity.
The conceptual advance here is that this derivation
uncovers hidden off-diagonal
correlations of the GL theory whose range is far
longer than that of the original superconducting correlator.
These hidden correlations reflect the collapse of the
LL structure brought about by the strong inter-LL mixing.
Second, when it comes to the theory of real
type-II superconductivity, the main
utility of this novel formulation is in the fact that
it describes the {\it finite-field} [FF] critical
behavior in contrast to the familiar zero-field [ZF]
critical point at $T=T_{c0}$. The strong fluctuation regime
of type-II superconductors for $T<T_{c0}$ and low fields
will be controlled by the proximity
to such a FF transition rather than the ZF one, contrary
to what is often assumed in the literature.\cite{schneider}
Even if the predicted FF transition itself
turns out to be hard to access in a
real experiment,\cite{first} the new
description introduced here should still be valuable in
providing a systematic approach to physical problems which up to
now were beyond analytical reach: The construction of thermodynamic
functions describing the ZF-FF crossover, the renormalization
of London vortex interactions by critical fluctuations, the
issue of the high-field versus the low-field thermodynamic scaling, etc.

I am grateful to Prof. M. Salamon
for asking important
questions, to Dr. I. F. Herbut, Prof. S. Teitel and
Prof. O. T. Valls for discussions,
and to Prof. T. C. Lubensky for
explaining to me the
subtleties of Ref. \cite{lubensky}.
This work has been supported in part by the NSF grant
DMR-9415549.


\newpage

\begin{figure}
\caption{A schematic representation of the critical region.
LVL -- London (s)vortex liquid; LVS -- London (s)vortex solid;
SCDW -- charge density-wave of Cooper pairs
(see Te{\v s}anovi{\' c} in Ref. [1]). The dotted line indicates
that $H$ is too high for the semiclassical approximation
to be reliable in determining $H_{\Phi}(T)$. The dashed
line is the mean-field $H_{c2}(T)$. The full line represents
the London (s)vortex solid-liquid melting transition
in the low-field regime while it separates the normal
state from the density-wave of Cooper pairs at high fields.
The low-field melting line is well-separated from
$H_{\Phi}(T)$ because the melting transition
takes place only for a rather large $\Gamma$
($\Gamma_{\rm melt}\sim 140$) and is well below the
nominal critical region (see Eq. (6) and the text above it).
The arrow on the $H$-axis indicates the crossover from
the high-field to the low-field regime of critical behavior,
$H_b\sim (\theta/16)(T/T_{c0})H_{c2}(0)\sim 1$ Tesla in HTS.
The inset shows regions of
`XY' and `$\Phi$' critical behavior
as described in text. The
dashed-dotted XY-$\Phi$ crossover line is
set by $\xi_{\Psi}(T,H=0)\sim\ell$.}
\label{fig1}
\end{figure}


\begin{references}

\bibitem{gl-lll}
G. J. Ruggeri and D. J. Thouless, J. Phys. F {\bf 6}, 2063 (1976);
E. Br\' ezin, D. R. Nelson, and A. Thiaville,
Phys. Rev. B {\bf 31}, 7124 (1985); M. A. Moore,
Phys. Rev. B {\bf 39}, 136 (1989);
E. Br{\' e}zin, A. Fujita, and S. Hikami, Phys. Rev. Lett.
{\bf 65}, 1949 (1990);
S. Ullah and A. T. Dorsey, Phys. Rev. Lett. {\bf 65}, 2066 (1990);
Z. Te\v{s}anovi\'{c} and L. Xing,
Phys. Rev. Lett. {\bf 67}, 2729 (1991);
Z. Te{\v s}anovi\'{c}, Phys. Rev. B
{\bf 44}, 12635 (1991); {\bf 46},
5884(E) (1992);
Y. Kato and N. Nagaosa, Phys. Rev. B {\bf 47}, 2932 (1993);
J. Hu and A. H. MacDonald, Phys. Rev. Lett. {\bf 71}, 432 (1993);
R. \v S\' a\v sik and D. Stroud, Phys. Rev. B {\bf 49}, 16074 (1994);
S. A. Ktitorov, B. N. Shalaev, and L. Jastrabik, Phys. Rev. B
{\bf 49}, 15248 (1994);
Z. Te\v {s}anovi\' {c},
Physica (Amsterdam) {\bf 220C}, 303 (1994).

\bibitem{vortex}
D. R. Nelson, Phys. Rev. Lett. {\bf 60}, 1973 (1988);
M. P. A. Fisher, Phys. Rev. Lett. {\bf 62}, 1415 (1989);
D. R. Nelson and H. S. Seung, Phys. Rev. B {\bf 39}, 9153 (1989);
M. V. Feigel'man, V. B. Geshkenbein, and
V. M. Vinokur,
JETP Lett. {\bf 52}, 546 (1990);
D. R. Nelson and P. Le Doussal, Phys. Rev. B {\bf 42},
10113 (1990);
D. S. Fisher, M. P. A. Fisher, and D. A. Huse, Phys. Rev. B
{\bf 43}, 130 (1991);
L. I. Glazman and A. E. Koshelev,
Phys. Rev. B  {\bf 43}, 2835 (1991);
L. N. Bulaevskii, M. Ledvij, and V. G. Kogan,
Phys. Rev. Lett. {\bf  68}, 3773 (1992);
G. Blatter and B. Ivlev,
{\it ibid.} {\bf 70}, 2621 (1993);
M. V. Feigel'man, V. B. Geshkenbein, L. B. Ioffe,
and A. I. Larkin, Phys. Rev. B {\bf 48}, 16641 (1994).

\bibitem{zt}
Z. Te{\v s}anovi\'{c}, Phys. Rev. B {\bf 44}, 12635 (1991); {\bf 46},
5884(E) (1992);

\bibitem{ta}
Z. Te\v sanovi\' c and A. V. Andreev,
Phys. Rev. B {\bf 49}, 4064 (1994).

\bibitem{ikeda}
The semiclassical description proposed
in this paper cannot be reached within the Gaussian
approximation or various perturbative schemes
applied to the original $F_{GL}$ (1):
R. Ikeda, T. Ohmi, T. Tsuneto, Phys. Rev. Lett. {\bf 67}, 3874 (1991);
J. Phys. Soc. Japan, {\bf 60}, 1051 (1991);
I. D. Lawrie, Phys. Rev. B {\bf 50}, 9456 (1994).
Such approaches are unable to produce
strong inter-LL correlations which are required in
the present description.

\bibitem{ocp}
A. Alastuey and B. Jancovici,
J. Physique {\bf 42}, 1 (1981),
and references therein.

\bibitem{schmid}
A. Schmid, Phys. Rev. {\bf 180}, 527 (1969);
H. Schmidt, Z. Phys. {\bf 216}, 336 (1968).

\bibitem{blatter}
G. Blatter, V. B. Geshkenbein, and A. I. Larkin,
Phys. Rev. Lett. {\bf 68}, 875 (1992); H. Hao
and J. R. Clem, Phys. Rev. B {\bf 46}, 5853 (1992).

\bibitem{gauge}
The gauge is fixed since ${\bf s}$ is purely
transverse in the xy-plane.

\bibitem{degennes}
P. G. DeGennes, Solid State Com. {\bf 10}, 753 (1972).

\bibitem{lubensky}
B. I. Halperin and T. C. Lubensky, Solid State Comm. {\bf 14},
997 (1974); B. I. Halperin, T. C. Lubensky
and S. Ma, Phys. Rev. Lett. {\bf 32}, 292 (1974);
T. C. Lubensky and J-H. Chen, Phys. Rev. B {\bf 17},
366 (1978).

\bibitem{first}
As shown in Ref. [11] the standard SHE also exhibits
a first-order transition, both within the mean-field
(compare with Eq. (6))
and the RG analysis. Note that, in our case, at very large
distances and/or very low fields, the
`real' electromagnetic field fluctuations
may have to be included as well since $\kappa^2$, while
large, is still finite. This will clearly modify
the FF critical behavior and may lead to the ${\bar\Phi}$-transition
becoming only a sharp crossover or a weak first
order transition involving no symmetry breaking.
The effect of these real electromagnetic field
fluctuations, while clearly of theoretical interest,
is of lesser practical significance in HTS
and other extremely type-II systems. In these materials
$\kappa^2$ is as large as 10$^3$--10$^4$ and the
magnetic field penetration depth in the critical
region will typically be longer than the
lengthscale over which the system can be considered
homogeneous. This is reminiscent of the
issue of smearing of the Kosterlitz-Thouless-Berezinskii (KTB)
transition in superconducting films by electromagnetic screening.
In practice, the effective magnetic field penetration depth
is often larger than the sample size and one can
observe the original $\kappa\to\infty$ KTB fluctuation behavior.

\bibitem{footxii}
Since our theory does not have the local gauge symmetry
of the SHE, such terms are generally permitted.
If $K^0_{\parallel,\perp}/m_s^2\sim\Lambda^2$,
sgauge field fluctuations will be
reduced throughout the critical
region. For discussion of related issues in the `pure'
SHE see M. Kiometzis, H. Kleinert, and A. M. J.
Schakel, Phys. Rev. Lett. {\bf 73}, 1975 (1994).

\bibitem{dasgupta}
C. Dasgupta and B. I. Halperin, Phys. Rev. Lett. {\bf 47},
1556 (1981).

\bibitem{bart}
J. Bartholomew, Phys. Rev. {\bf B28}, 5378 (1983).

\bibitem{leo}
L. Radzihovsky, Europhys. Lett. {\bf 29}, 227 (1995).

\bibitem{ott}
R. Jin, A. Schilling and H. R. Ott, Phys. Rev. B {\bf 49},
9218 (1994).

\bibitem{cv}
S. E. Inderhees, M. B. Salamon, J. P. Rice, and
D. M. Ginsberg, Phys. Rev. Lett. {\bf 66}, 232 (1992).

\bibitem{pierson}
S. W. Pierson, Ph.D. Thesis, University of Minnesota, 1993.

\bibitem{teitel}
Y. H. Li and S. Teitel,  Phys. Rev. B {\bf 49}, 4136 (1994);
S. Teitel, private communication.  More precisely,
these authors observe that,
at some well defined temperature $T_{cz}$, an interconnected
tangle of wandering field-induced vortices and thermally-induced
vortex rings percolates through the system in the plane
perpendicular to the applied field. At the same temperature,
the helicity modulus along the field direction, which was found
to be finite below $T_{cz}$, goes continuously to zero. This
is just what one expects at the $\bar\Phi$-transition of
our model: As one approaches $T_{cz}$  from below, large thermally
induced vortex loops in $\bar\Phi$ proliferate across the system.
Such vortex excitations are confined within
a flux tube of the fictitious gauge field.  This flux tube
represents the screening cloud of field- and thermally-induced
vorticity and corresponds to the above ``interconnected tangle"
observed in the Monte Carlo simulations of Li and Teitel. The
cross-sectional area of such a tangle is given by the product of
pertinent penetration depths of our fictitious electrodynamics.
The helicity modulus along the field directly measures the
formation of a {\it uniform} $\bar\Phi$-condensate. Thus, their $T_{cz}$
can be identified with $T_{\Phi}(H)$. The resistivity
anomalies along the field observed by H. Safar {\it et al.},
Phys. Rev. Lett. {\bf 72}, 1272 (1994), are presumably also related
to $T_{\Phi}(H)$.  In this context, it should be noted that the label
London vortex liquid (LVL) below $T_{\Phi}(H)$ (Fig. 1)
indicates a translationally invariant phase in which
the long wavelength density-density interaction of
the fluctuating vorticity
has a characteristic London form, the strength of which
is given by the superfluid density of the $\bar\Phi$-condensate.
This phase has
strong thermal fluctuations throughout the critical region
which makes it impossible to locally separate
field-induced from thermally-induced vortex excitations.
Thus, the LVL phase (Fig. 1) should be clearly distinguished
from the often used London line liquid description (Ref. 1)
in which only field-induced vortices are considered
and all other fluctuations are ignored.  Such description
is applicable only at temperatures well below $T_{\Phi}(H)$.

\bibitem{schneider}
T. Schneider and H. Keller,
Int. J. Mod. Phys. B {\bf 8}, 487 (1994).


\end{references}
\end{document}